\documentclass[superscriptaddress,nofootinbib,preprintnumbers,floatfix,twocolumn,prd]{revtex4}

\usepackage{amsmath}
\usepackage{amssymb}
\usepackage{graphicx}
\usepackage{slashed}
\usepackage{floatrow}
\usepackage{tikz}
\usepackage{color}
\def\nn{\nonumber \\ }
\usepackage{enumerate}

\def\vev#1{\left\langle #1 \right \rangle}
\def\abs#1{\left| #1 \right |}
\def\rd{ {\rm d}}
\newcommand{\at}{\makeatletter @\makeatother}
\def\mgv{{\sc MadGraph5\_aMC\at NLO}}
\def\scetew{$\text{SCET}_{\text{EW}}$}

\usetikzlibrary{arrows,decorations.pathmorphing,backgrounds,positioning,fit,petri,automata,shadows,calendar,mindmap,decorations.markings,calc}

\def\lL{\mathsf{L}}
\def\lM{\mathsf{L_M}}
\def\lQ{\mathsf{L_Q}}

\begin{document}

\preprint{MITP/14-063}

\title{Non-cancellation of electroweak logarithms in high-energy scattering}

\author{Aneesh V.~Manohar}
\affiliation{Department of Physics, University of California at San Diego, La Jolla, CA 92093, USA}

\author{Brian Shotwell}
\affiliation{Department of Physics, University of California at San Diego, La Jolla, CA 92093, USA}

\author{Christian W.~Bauer}
\affiliation{Ernest Orlando Lawrence Berkeley National Laboratory, University of California, Berkeley, CA 94720, USA}

\author{Sascha Turczyk}
\affiliation{PRISMA Cluster of Excellence \& Mainz Institut for Theoretical Physics, Johannes Gutenberg University, 55099 Mainz, Germany}

\date{\today}

\begin{abstract}

We study electroweak Sudakov corrections in high energy scattering, and the cancellation between real and virtual Sudakov corrections. Numerical results are given for the case of heavy quark production by gluon collisions involving the rates $gg \to t \bar t, b \bar b, t \bar b W, t \bar t Z, b \bar b Z, t \bar t H, b \bar b H$. Gauge boson virtual corrections are related to real transverse gauge boson emission, and Higgs virtual corrections to Higgs and longitudinal gauge boson emission. At the LHC, electroweak corrections become important in the TeV regime. At the proposed 100\,TeV collider, electroweak interactions enter a new regime, where the corrections are very large and need to be resummed.

\end{abstract}

\maketitle

\section{Introduction}\label{sec:intro}

Electroweak corrections grow with energy due to the presence of Sudakov double logarithms $\alpha_W \ln^2 s/M_W^2$, and are already relevant for LHC analyses with invariant masses in the TeV region. The corrections arise because of soft and collinear infrared divergences from the emission of electroweak bosons. The infrared singularities are cutoff by the gauge boson mass, and lead to finite $\alpha_W \ln^2 s/M_W^2$ corrections. Unlike in QCD, the electroweak logarithms do not cancel even for totally inclusive processes, because the initial states are not electroweak singlets~\cite{ccc,ciafaloni1,ciafaloni2}. 

In this paper, we discuss the cancellation (or lack thereof) between real and virtual corrections. We will use $ g g \to t \bar t, b \bar b$ as an explicit numerical example. In this process, the initial state \emph{is} an electroweak singlet, so the total cross section does not contain $\alpha_W \ln^2 s/M_W^2$ corrections. This allows us to compare the electroweak corrections in this process to the more familiar case of $\alpha_s$ corrections to the $R$ ratio for $e^+ e^- \to \text{hadrons}$. Even though electroweak Sudakov corrections cancel for the total cross section, they do not cancel for  interesting experimentally measured rates, and are around 10\% for invariant masses of $\sim 2$\,TeV. Some earlier work related to our paper can be found in Refs.~\cite{Ciafaloni:2006qu,Baur:2006sn,Bell:2010gi,Frixione:2014qaa}. Electroweak corrections to processes involving electroweak-charged initial states, such as Drell-Yan production, $q \overline q \to WW$, or $q \overline q \to t \overline t$, are larger than 
for $gg \to t \overline t$. 

At present, omitted electroweak corrections are the largest error in many LHC cross section calculations, and are more important than higher order QCD corrections. Furthermore, the resummed electroweak corrections to \emph{all} hard scattering processes at NLL order are known explicitly~\cite{Chiu:2009mg,Chiu:2009ft,Fuhrer:2010eu}, and have a very simple form, so they can be incorporated into LHC cross section calculations. Recently, there has been interest in building a hadron collider with an energy of around 100\,TeV. For such a machine, electroweak corrections are no longer small, and resummed corrections must be included to get reliable cross sections. The numerical plots in this paper go out to $\sqrt{\hat s} = 30$\,TeV to emphasize the importance of electroweak corrections at future machines.

We will make one simplification in this paper, by computing electroweak corrections in a pure $SU(2)_W$ gauge theory, neglecting the $U(1)$ part. The reason is that in the Standard Model (SM), after spontaneous symmetry breaking, there is a massless photon. Electromagnetic corrections produce infrared divergences which are not regularized by a gauge boson mass. Instead they have to be treated by defining infrared safe observables, as done for QCD. Initial state infrared corrections can be absorbed into the parton distribution functions (PDFs). To implement this consistently requires electromagnetic corrections to be included in the parton evolution equations. These additional complications are separate from the main point of the paper, and can be avoided by using the $SU(2)_W$ theory.

The numerical results will be given for an $SU(2)_W$ gauge theory with $\alpha_W$ equal to the Standard Model value $\alpha/\sin^2 \theta_W$. We will treat $W^{1,2}$ as the SM $W$ bosons, and $W^3$ as the SM $Z^0$, and use the notation $\lL \equiv \ln s/M_W^2$.

The structure of electroweak corrections is discussed in Sec.~\ref{sec:ew}, and a summary of the \scetew\ results for computing these is given in Appendix~\ref{sec:scet}. The  cancellation of real and virtual electroweak corrections is discussed in Sec.~\ref{sec:cancel} for an example where one can do the full computation analytically, and Sec.~\ref{sec:tt} discusses the cancellation for heavy quark production, where the rates have to be computed numerically. Some subtleties for an unstable $t$-quark are discussed in Sec.~\ref{sec:47}. The implications of electroweak corrections for experimental measurements is given in Sec.~\ref{sec:conclusions}.

\section{Electroweak Logarithms}\label{sec:ew}

Electroweak radiative corrections have a typical size of order $\alpha_W/\pi \sim 0.01$. However, in some cases, the radiative corrections have a Sudakov double logarithm, $(\alpha_W/\pi)\lL^2 $, and  become important. The regime where this happens is high energy, $s \gg M_W^2$, where one can apply soft-collinear effective theory (SCET)~\cite{Bauer:2000ew,Bauer:2000yr,Bauer:2001ct,Bauer:2001yt}. The electroweak version of SCET (\scetew) was developed in a series of papers Refs.~\cite{Chiu:2007yn,Chiu:2007dg,Chiu:2008vv,Chiu:2009yz,Chiu:2009yx,Chiu:2009mg,Chiu:2009ft,Fuhrer:2010eu,Fuhrer:2010vi}, and has  important differences from the QCD case, namely the presence of a broken gauge symmetry, massive gauge bosons, and multiple mass scales $M_Z$, $M_W$, $M_H$ and $m_t$. The effective theory is a systematic expansion in $M_W^2/s$, and at leading order, all $(M_W^2/s)^n$ power corrections are omitted. The neglect of these power corrections greatly simplifies the computation, and the electroweak corrections have 
an elegant universal form. We are in the lucky situation where the theory simplifies in the regime where the electroweak corrections are important. Electroweak corrections have been computed by many groups by other methods~\cite{ccc,ciafaloni1,ciafaloni2,fadin,kps,fkps,jkps,jkps4,beccaria,dp1,dp2,hori,beenakker,dmp,pozzorini,js,melles1,melles2,melles3,Denner:2006jr,kuhnW,Denner:2008yn}.

It is instructive to compare the SCET result with the vastly more difficult conventional fixed order approach to computing electroweak corrections. At fixed order one gets an expansion $\sum_{n,r} c_{n,r}\alpha_W^n \lL^r$ with $r \le 2n$, which breaks down at high energies. Furthermore, one has to do a very difficult multi-scale computation (with scales $s$, $M_Z$, $M_H$, $m_t$) for each new process being considered. The fixed order results are available only for a few cases, and often with the approximation that $M_W=M_Z=M_H$. In contrast, the SCET result, Eq.~(\ref{m}) has a simple form where all the pieces are known, so each new process can be computed by multiplying the appropriate factors, which are all known in closed form. The reason the fixed order calculation is much harder, of course, is that it includes the $M_Z^2/s$ power corrections, which then have to be expanded out. The $M_Z^2/s$  power corrections are negligible in the region where  electroweak corrections are large and experimentally 
important. We summarize the \scetew\ results in Appendix~\ref{sec:scet}. More details can be found in Refs.~\cite{Chiu:2007yn,Chiu:2007dg,Chiu:2008vv,Chiu:2009yz,Chiu:2009yx,Chiu:2009mg,Chiu:2009ft,Fuhrer:2010eu}.

An explicit numerical analysis comparing fixed order and \scetew\ results is given in Sec.~\ref{sec:cancel}.

\section{Cancellation of Real and Virtual Corrections}\label{sec:cancel}

Recall the familiar example of the total cross section for $e^+ e^- \to \text{hadrons}$, which has an expansion in $\alpha_s(Q^2)$, with no large logarithms. At one-loop, the virtual correction to $e^+ e^- \to q \overline q$ is infrared divergent, as is the $e^+ e^- \to q \overline q g$ real radiation rate, but the sum of the two is infrared finite, and gives the correction to the ratio of the $e^+ e^-$ total cross section to its tree-level value,  $R=1+\alpha_s/\pi$.

The electroweak corrections to $gg \to q \overline q$ have a similar cancellation. Rather than study this process, we  first start with the simpler case of $J \to q \overline q$, where $J^\mu=\overline q \gamma^\mu P_L q$ is an external gauge invariant current that produces the  doublet $q_L=(t,b)_L$, where we treat $t$ and $b$ as massless quarks. The main reason for doing this is to avoid complicated phase space integrals for real radiation, and fermion mass effects, and because it is closely related to the familiar QCD case of $R$. The $ gg \to q \overline q$ case with $q_L=(t,b)_L$will be studied numerically in Sec.~\ref{sec:tt}.

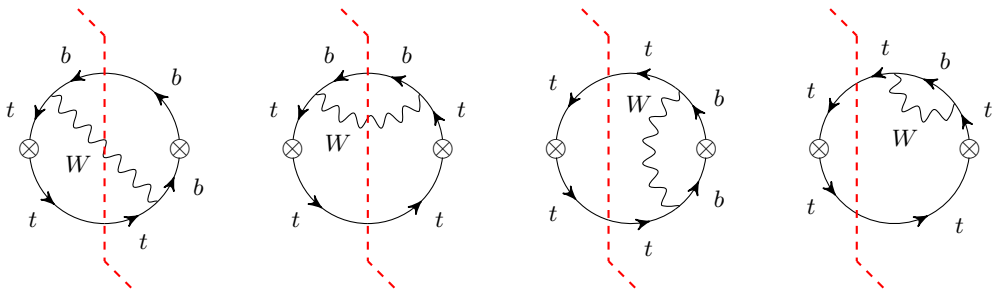
\begin{figure*}
\ffigbox{ 
\begin{tikzpicture}

\draw (0-1,0) node [align=center] {\scalebox{1.2}{$\otimes$}};
\draw (0+1,0) node [align=center] {\scalebox{1.2}{$\otimes$}};
\draw[postaction=decorate] [decoration={markings, mark=at position 0.50 with {\arrow[scale=1.5]{stealth'}};}] (0,0)+(7:1) arc (7:90:1);
\draw[postaction=decorate] [decoration={markings, mark=at position 0.60 with {\arrow[scale=1.5]{stealth'}};}] (0,0)+(90:1) arc (90:135:1);
\draw[postaction=decorate] [decoration={markings, mark=at position 0.60 with {\arrow[scale=1.5]{stealth'}};}] (0,0)+(135:1) arc (135:173:1);
\draw[postaction=decorate] [decoration={markings, mark=at position 0.50 with {\arrow[scale=1.5]{stealth'}};}] (0,0)+(187:1) arc (187:270:1);
\draw[postaction=decorate] [decoration={markings, mark=at position 0.60 with {\arrow[scale=1.5]{stealth'}};}] (0,0)+(0,-1) arc (270:315:1);
\draw[postaction=decorate] [decoration={markings, mark=at position 0.60 with {\arrow[scale=1.5]{stealth'}};}] (0,0)+(315:1) arc (315:353:1);
\draw[decorate,decoration={snake,segment length=7.95pt}] (0,0)+(-45:1) -- ++(135:1);
\draw[dashed,red,thick] (0,1.5)+(135:0.5) -- (0,1.5);
\draw[dashed,red,thick] (0,1.5) -- (0,-1.5);
\draw[dashed,red,thick] (0,-1.5)+(315:0.5) -- (0,-1.5);
\draw (0,0)+(45:1.35) node [align=center] {$b$};
\draw (0,0)+(112.5:1.35) node [align=center] {$b$};
\draw (0,0)+(157.5:1.35) node [align=center] {$t$};
\draw (0,0)+(225:1.35) node [align=center] {$t$};
\draw (0,0)+(292.5:1.35) node [align=center] {$t$};
\draw (0,0)+(337.5:1.35) node [align=center] {$b$};
\draw (0,0)+(210:0.4) node [align=center] {$W$};

\draw (3.5-1,0) node [align=center] {\scalebox{1.2}{$\otimes$}};
\draw (3.5+1,0) node [align=center] {\scalebox{1.2}{$\otimes$}};
\draw[postaction=decorate] [decoration={markings, mark=at position 0.60 with {\arrow[scale=1.5]{stealth'}};}] (3.5,0)+(7:1) arc (7:45:1);
\draw[postaction=decorate] [decoration={markings, mark=at position 0.60 with {\arrow[scale=1.5]{stealth'}};}] (3.5,0)+(45:1) arc (45:90:1);
\draw[postaction=decorate] [decoration={markings, mark=at position 0.60 with {\arrow[scale=1.5]{stealth'}};}] (3.5,0)+(90:1) arc (90:135:1);
\draw[postaction=decorate] [decoration={markings, mark=at position 0.60 with {\arrow[scale=1.5]{stealth'}};}] (3.5,0)+(135:1) arc (135:173:1);
\draw[postaction=decorate] [decoration={markings, mark=at position 0.55 with {\arrow[scale=1.5]{stealth'}};}] (3.5,0)+(187:1) arc (187:270:1);
\draw[postaction=decorate] [decoration={markings, mark=at position 0.55 with {\arrow[scale=1.5]{stealth'}};}] (3.5,0)+(0,-1) arc (270:353:1);
\draw[decorate,decoration={snake,segment length=7.25pt}] (3.5,1.3)+(220:0.9) arc (220:320:0.9);
\draw[dashed,red,thick] (3.5,1.5)+(135:0.5) -- (3.5,1.5);
\draw[dashed,red,thick] (3.5,1.5) -- (3.5,-1.5);
\draw[dashed,red,thick] (3.5,-1.5)+(315:0.5) -- (3.5,-1.5);
\draw (3.5,0)+(22.5:1.35) node [align=center] {$t$};
\draw (3.5,0)+(67.5:1.35) node [align=center] {$b$};
\draw (3.5,0)+(112.5:1.35) node [align=center] {$b$};
\draw (3.5,0)+(157.5:1.35) node [align=center] {$t$};
\draw (3.5,0)+(225:1.35) node [align=center] {$t$};
\draw (3.5,0)+(315:1.35) node [align=center] {$t$};
\draw (3.5,0)+(170:0.4) node [align=center] {$W$};

\draw (7-1,0) node [align=center] {\scalebox{1.2}{$\otimes$}};
\draw (7+1,0) node [align=center] {\scalebox{1.2}{$\otimes$}};
\draw[postaction=decorate] [decoration={markings, mark=at position 0.60 with {\arrow[scale=1.5]{stealth'}};}] (7,0)+(7:1) arc (7:54:1);
\draw[postaction=decorate] [decoration={markings, mark=at position 0.60 with {\arrow[scale=1.5]{stealth'}};}] (7,0)+(54:1) arc (54:107.46:1);
\draw[postaction=decorate] [decoration={markings, mark=at position 0.60 with {\arrow[scale=1.5]{stealth'}};}] (7,0)+(107.46:1) arc (107.46:173:1);
\draw[postaction=decorate] [decoration={markings, mark=at position 0.55 with {\arrow[scale=1.5]{stealth'}};}] (7,0)+(187:1) arc (187:252.54:1);
\draw[postaction=decorate] [decoration={markings, mark=at position 0.60 with {\arrow[scale=1.5]{stealth'}};}] (7,0)+(252.54:1) arc (252.54:306:1);
\draw[postaction=decorate] [decoration={markings, mark=at position 0.60 with {\arrow[scale=1.5]{stealth'}};}] (7,0)+(306:1) arc (306:353:1);
\draw[decorate,decoration={snake,segment length=8.1pt}] (7+1.29,0)+(130:1) arc (130:230:1);
\draw[dashed,red,thick] (7-.3,1.5)+(135:0.5) -- (7-.3,1.5);
\draw[dashed,red,thick] (7-.3,1.5) -- (7-.3,-1.5);
\draw[dashed,red,thick] (7-.3,-1.5)+(315:0.5) -- (7-.3,-1.5);
\draw (7,0)+(30:1.35) node [align=center] {$b$};
\draw (7,0)+(80:1.35) node [align=center] {$t$};
\draw (7,0)+(140:1.35) node [align=center] {$t$};
\draw (7,0)+(220:1.35) node [align=center] {$t$};
\draw (7,0)+(280:1.35) node [align=center] {$t$};
\draw (7,0)+(330:1.35) node [align=center] {$b$};
\draw (7,0)+(80:0.6) node [align=center] {$W$};

\draw (10.5-1,0) node [align=center] {\scalebox{1.2}{$\otimes$}};
\draw (10.5+1,0) node [align=center] {\scalebox{1.2}{$\otimes$}};
\draw[postaction=decorate] [decoration={markings, mark=at position 0.60 with {\arrow[scale=1.5]{stealth'}};}] (10.5,0)+(7:1) arc (7:40:1);
\draw[postaction=decorate] [decoration={markings, mark=at position 0.55 with {\arrow[scale=1.5]{stealth'}};}] (10.5,0)+(40:1) arc (40:90:1);
\draw[postaction=decorate] [decoration={markings, mark=at position 0.60 with {\arrow[scale=1.5]{stealth'}};}] (10.5,0)+(90:1) arc (90:120:1);
\draw[postaction=decorate] [decoration={markings, mark=at position 0.60 with {\arrow[scale=1.5]{stealth'}};}] (10.5,0)+(120:1) arc (120:173:1);
\draw[postaction=decorate] [decoration={markings, mark=at position 0.60 with {\arrow[scale=1.5]{stealth'}};}] (10.5,0)+(187:1) arc (187:240:1);
\draw[postaction=decorate] [decoration={markings, mark=at position 0.50 with {\arrow[scale=1.5]{stealth'}};}] (10.5,0)+(240:1) arc (240:353:1);
\draw[decorate,decoration={snake,segment length=7.62pt}] (10.5,1) arc (180:307:0.5);
\draw[dashed,red,thick] (10.5-.5,1.5)+(135:0.5) -- (10.5-.5,1.5);
\draw[dashed,red,thick] (10.5-.5,1.5) -- (10.5-.5,-1.5);
\draw[dashed,red,thick] (10.5-.5,-1.5)+(315:0.5) -- (10.5-.5,-1.5);
\draw (10.5,0)+(22.5:1.35) node [align=center] {$t$};
\draw (10.5,0)+(60:1.35) node [align=center] {$b$};
\draw (10.5,0)+(95:1.35) node [align=center] {$t$};
\draw (10.5,0)+(145:1.35) node [align=center] {$t$};
\draw (10.5,0)+(215:1.35) node [align=center] {$t$};
\draw (10.5,0)+(310:1.35) node [align=center] {$t$};
\draw (10.5,0)+(45:0.2) node [align=center] {$W$};

\end{tikzpicture}
}{\caption{\label{fig:cut} Graphs contributing to the $\alpha_W$ correction to the $J \to q \bar q$ rate.}}
\end{figure*}

The total cross section for $J \to q \overline q$ can be written as the imaginary part of the vacuum bubbles $\Pi(Q^2)$ in Fig.~\ref{fig:cut}. $\Pi(Q^2)$ at Euclidean $Q^2$ is infrared finite. Thus the analytic continuation to Minkowski space is also infrared finite, and the sum of the real and virtual rates, which is equal to the imaginary part of $\Pi(Q^2)$, is infrared finite.

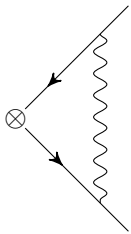
\begin{figure}
\ffigbox{ 
\begin{tikzpicture}[scale=1.5]

\draw (6,0) node [align=center] {\scalebox{1.2}{$\otimes$}};
\draw[postaction=decorate] [decoration={markings, mark=at position 0.60 with {\arrowreversed[scale=1.5]{stealth'}};}]  (6,0)+(45:0.12) -- (6.5,0.5);
\draw[postaction=decorate] [decoration={markings, mark=at position 0.80 with {\arrow[scale=1.5]{stealth'}};}]  (6,0)+(315:0.12) -- (6.5,-0.5);
\draw (6.5,0.5) -- (7,1);
\draw (6.5,-0.5) -- (7,-1);
\draw[decorate,decoration={snake,segment length=8.25pt}] (6.75,0.75) -- (6.75,-0.75);

\end{tikzpicture}
}{\caption{\label{fig:virtual} Virtual correction to $J \to q \overline q$.}}
\end{figure}

The virtual correction to $J \to q \overline q$ is given by the graph in Fig.~\ref{fig:virtual} and wave-function graphs, 
which gives the vertex form-factor
\begin{align}
F_V &=1+\frac{C_F\alpha_W }{4\pi} \Bigl\{-\frac72+2 \widetilde r -(3-2 \widetilde r) \ln \widetilde r \nn
& +\left(1-\widetilde r\right)^2 \bigl[2 \text{Li}_2(\widetilde r)-\ln^2 \widetilde r+2 \ln \widetilde r \ln (1-\widetilde r)-\frac{2 \pi ^2}{3} \bigr]\Bigr\}\nn
\end{align}
at Euclidean momentum transfer $q^2=-Q^2<0$, with
\begin{align}
\widetilde r &= \frac{M_W^2}{Q^2},
\end{align}
where $M_W$ is the gauge boson mass.\footnote{Eq.~(12) of Ref.~\cite{Chiu:2009mg} is incorrect near threshold.} Analytically continuing to time-like $q^2=s>0$,
\begin{align}
r &= \frac{M^2}{s} =-\widetilde r, & \ln \widetilde r &= \ln r - i \pi,
\end{align}
gives
\begin{align}
F_V &=1+\frac{C_F\alpha_W }{4\pi} \Bigl\{-\frac72-2  r -(3+2  r) \ln  r \nn
& +\left(1+ r\right)^2 \bigl[2 \text{Li}_2(- r)-\ln^2  r+2 \ln  r \ln (1+ r)+\frac{ \pi ^2}{3} \bigr]\nn
&+  i \pi \bigl[  (3+2  r)+2\left(1+ r\right)^2 \left(\ln r - \ln (1+r) \right)\bigr]\Bigr\}
\end{align}
which for $s \to \infty$ is
\begin{align}
F_V &=1+\frac{C_F\alpha_W }{4\pi} \times \nn
&\Bigl\{-\ln^2 r -3 \ln r +\frac{\pi^2}{3} -\frac{7}{2} + i \pi \left(2\ln r+3 \right) \Bigr\}
\label{FVexp}
\end{align}
The \scetew\ computation gives radiative corrections to the $J q \overline q $ operator neglecting $M^2/s$ power corrections, and gives  precisely Eq.~(\ref{FVexp}), when expanded out to order $\alpha_W$~\cite{Chiu:2007yn}. 

The one-loop virtual correction to the $J q\overline q $ cross section is (neglecting power corrections)
\begin{align}
\sigma_V &= \sigma_0 \left[\abs{F_V}^2-1\right] \nn
&= \sigma_0 \frac{C_F\alpha_W }{2\pi} \Bigl\{-\ln^2 r -3 \ln r +\frac{\pi^2}{3} -\frac{7}{2} \Bigr\}
\label{sigv}
\end{align}
where $\sigma_0$ is the tree-level cross section. The $-\ln^2 r$ and $-3 \ln r$ terms lead to large corrections at high energy.

The real radiation $J \to q \overline q W$ arises from the graphs in Fig.~\ref{fig:real}, and is
\begin{figure}
\ffigbox{ 
\begin{tikzpicture}[scale=1.5]

\draw (2,0) node [align=center] {\scalebox{1.2}{$\otimes$}};
\draw (2,0)+(45:0.12) -- (2.3,0.3);
\draw[postaction=decorate] [decoration={markings, mark=at position 0.50 with {\arrowreversed[scale=1.5]{stealth'}};}]  (2.3,0.3) -- (3,1);
\draw[decorate,decoration={snake,segment length=8pt}] (2.3,0.3) -- +(341.565:1.02);
\draw[postaction=decorate] [decoration={markings, mark=at position 0.60 with {\arrow[scale=1.5]{stealth'}};}]  (2,0)+(315:0.12) -- (3,-1);

\draw (4,0) node [align=center] {\scalebox{1.2}{$\otimes$}};
\draw[postaction=decorate] [decoration={markings, mark=at position 0.50 with {\arrowreversed[scale=1.5]{stealth'}};}]  (4,0)+(45:0.12) -- (5,1);
\draw (4,0)+(315:0.12) -- (4.3,-0.3);
\draw[decorate,decoration={snake,segment length=8pt}] (4.3,-0.3) -- +(18.435:1.02);
\draw[postaction=decorate] [decoration={markings, mark=at position 0.65 with {\arrow[scale=1.5]{stealth'}};}]  (4.3,-0.3) -- (5,-1);

\end{tikzpicture}
}{\caption{\label{fig:real} Real radiation from $J \to q \overline q W$.}}
\end{figure}
\begin{align}
\sigma_R 
&=\frac{C_F \alpha_W}{2\pi}  \sigma_0\biggl\{5(1-r^2)+ (3+4r+3r^2)\ln r\nn
&+(1+r)^2 \Bigl[\ln^2 r- 4\ln r \ln(1+r) - 4\,  \text{Li}_2 \left(-r\right)- \frac{\pi^2}{3}\Bigr]  \biggr\}.
\end{align}
Expanding in $r$ gives
\begin{align}
\sigma_R 
&=\frac{C_F \alpha_W}{2\pi}  \sigma_0\biggl\{ \ln^2r+3\ln r -\frac{\pi^2}{3}+5+\ldots  \biggr\}.
\label{sigr}
\end{align}

The total radiative correction is
\begin{align}
\sigma_T &=\sigma_R + \sigma_V\nn
&=\frac{C_F \alpha_W}{2\pi}  \sigma_0\biggl\{\frac32-2 r-5 r^2+ (2+3r)r \ln r\nn
&-2(1+r)^2\left[\ln r \ln(1+r)+  \text{Li}_2 \left(-r\right)\right]   \biggr\}
\end{align}
and as $r \to 0$ gives
\begin{align}
\sigma_T &=\frac{3C_F \alpha_W}{4\pi}  \sigma_0\,.
\label{R}
\end{align}
The $\ln^2 r$ and $\ln r$ terms cancel between $\sigma_{R,V}$. The 
correction to $R$ in QCD is given by Eq.~(\ref{R}) with the replacement $\alpha_W \to \alpha_s$ and $C_F \to 4/3$.

\begin{figure}
\includegraphics[width=8cm]{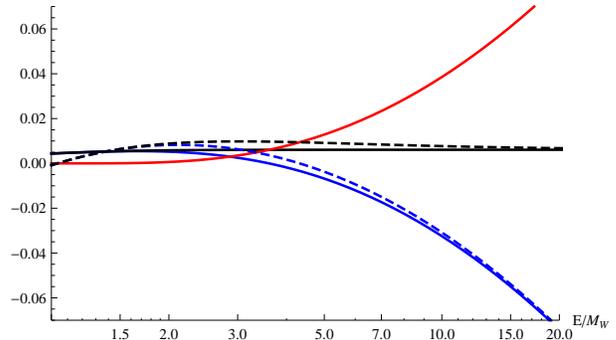}
\caption{\label{fig:fV} Plot of the real and virtual corrections to $J \to q \bar q$. Plotted are the exact virtual correction (solid blue), the virtual corrections using \scetew\ (dashed blue), real radiation (red), exact total rate (black) and the total rate using the \scetew\ virtual correction (dashed black).}
\end{figure}

The real and virtual corrections are shown in Fig.~\ref{fig:fV}. Also shown is the virtual correction computed using the \scetew\ result of Eq.~(\ref{m}).
The \scetew\ and exact calculations for the virtual correction have only very small differences, which are below 1\% for $E > 2M_W \sim 160$\, GeV, and $<0.5\%$ by 400\,GeV, whereas the real and virtual corrections each exceed 5\% by the time $E > 15 M_W \sim 1.2$\, TeV. This shows that in the regime where the electroweak corrections are relevant at the LHC, the \scetew\ computation is sufficiently accurate. The figure also shows that the large real and virtual electroweak corrections cancel in the total cross section.

The above calculation demonstrates the usual cancellation of the $\lL^2$ and $\lL$ terms between real and virtual graphs for the total cross section summed over all final states. This cancellation is not guaranteed to hold if the cross section is modified by restrictions on the final state. One can impose phase space restrictions on the kinematics of the emitted gauge boson. Consequences of doing so were studied in detail in Ref.~\cite{Bell:2010gi}, and lead to incomplete cancellation of the logarithms if the phase space cuts restrict the soft or collinear radiation. One can also investigate the possibility that, because electroweak charge is an experimental observable, one can separate the total cross section ($J \to t \overline t$, $b \overline b$, $t \overline t Z$, $b \overline b Z$, $t \overline b W^-$, $b \overline t W^+$) into sub-processes tagged by the final state particles, without restricting phase space. This is useful because the different channels have  different 
experimental signatures, and are often measured separately~\cite{ Khachatryan:2014ewa}. The second possibility is studied below, and is complementary to the non-cancellation of logarithmic terms due to phase space restrictions, and due to electroweak non-singlet initial states~\cite{ccc,ciafaloni1,ciafaloni2}.

The real and virtual cross sections are modified if one does not sum over all final states. In the simple example we are considering with degenerate fermions and bosons, the only change is that Eqs.~(\ref{sigv},\ref{sigr}) are modified by the replacement of the group theory factor $N C_F$ ($N=2$) by $G_{V}$ and $G_{R}$, which need not be equal, so that the total cross section
\begin{align}
\sigma_T
&=\frac{\alpha_W}{2\pi }\left(G_R-G_V \right) \widehat \sigma_0\biggl\{ \ln^2r+3\ln r +\ldots  \biggr\}
\label{sigt}
\end{align}
can have large corrections at high energy. The dependence of the cross section on $\ln^2 r + 3\ln r$ is characteristic of the IR structure of a vector current~\cite{Manohar:2003vb}.

To study this non-cancellation, we tabulate the group theory factors $G_{V,R}$ in Table~\ref{tab:group} for some possible choices of final state, for an $SU(N)$ gauge theory. In Eq.~(\ref{sigt}), $\sigma_0=N \widehat \sigma_0$ is the total tree-level rate, so that $\hat \sigma_0$ is $N$-independent. The different cases are:
\begin{enumerate}
    \item Any fermion with or without any gauge bosons, i.e.\ the full inclusive rate.
    \item Any fermion but no gauge boson, e.g. $t \bar t$, $b \bar b$, but not $t \bar t Z$, $b \overline b Z$, $t \bar b W^-$, 
$b \bar t W^+$.
    \item Specify one fermion with or without any gauge bosons, e.g.\ $t + X$, with $X=\bar t$, $\bar t Z$, $\bar b W^-$.
    \item Specify one fermion and no gauge bosons, e.g.\ $t + X$, with $X=\bar t$.
     \item Specify both fermions (labeled by $i,j$) with or without any gauge bosons, e.g. $i=j=1$ is $t \overline t X$, $i=1,j=2$ is $t \overline b X$, etc.
    \item Specify both fermions and require no gauge bosons. Same as the previous case but $X$ cannot contain gauge bosons.
 \end{enumerate}

One can see that for cases 1 and 3, the logarithmic terms are absent, while for all other cases, the logarithms survive and give rise to large corrections at high energies. 

\renewcommand{\arraystretch}{1.5}
\begin{table}
\begin{eqnarray*}
\begin{array}{|c|c|c|c|}
\hline
\text{ Case } & \ G_{R} \ & \ G_{V} \ & \ G_{R} - G_{V} \ \\
\hline
1 & N C_F & N C_F & 0 \\
2 & 0 & N C_F & - NC_F \\
3 & C_F & C_F & 0  \\
4 & 0 & C_F & -C_F \\
5 & \frac12  - \frac1{2N} \delta_{ij} & C_F \delta_{ij} & \frac12 - \frac{N}2 \delta_{ij} \\
6 & 0 & C_F \delta_{ij} & -C_F \delta_{ij} \\ 
%
%
%
\hline
\end{array}
\end{eqnarray*}
\caption{\label{tab:group} Group theory factors for real and virtual emission for an $SU(N)$ gauge theory. $C_F=(N^2-1)/(2N)$. The different cases are described in the text.}
\end{table}

\section{Heavy Quark Production}\label{sec:tt}

\begin{figure}
\includegraphics[width=2.5cm,angle=180]{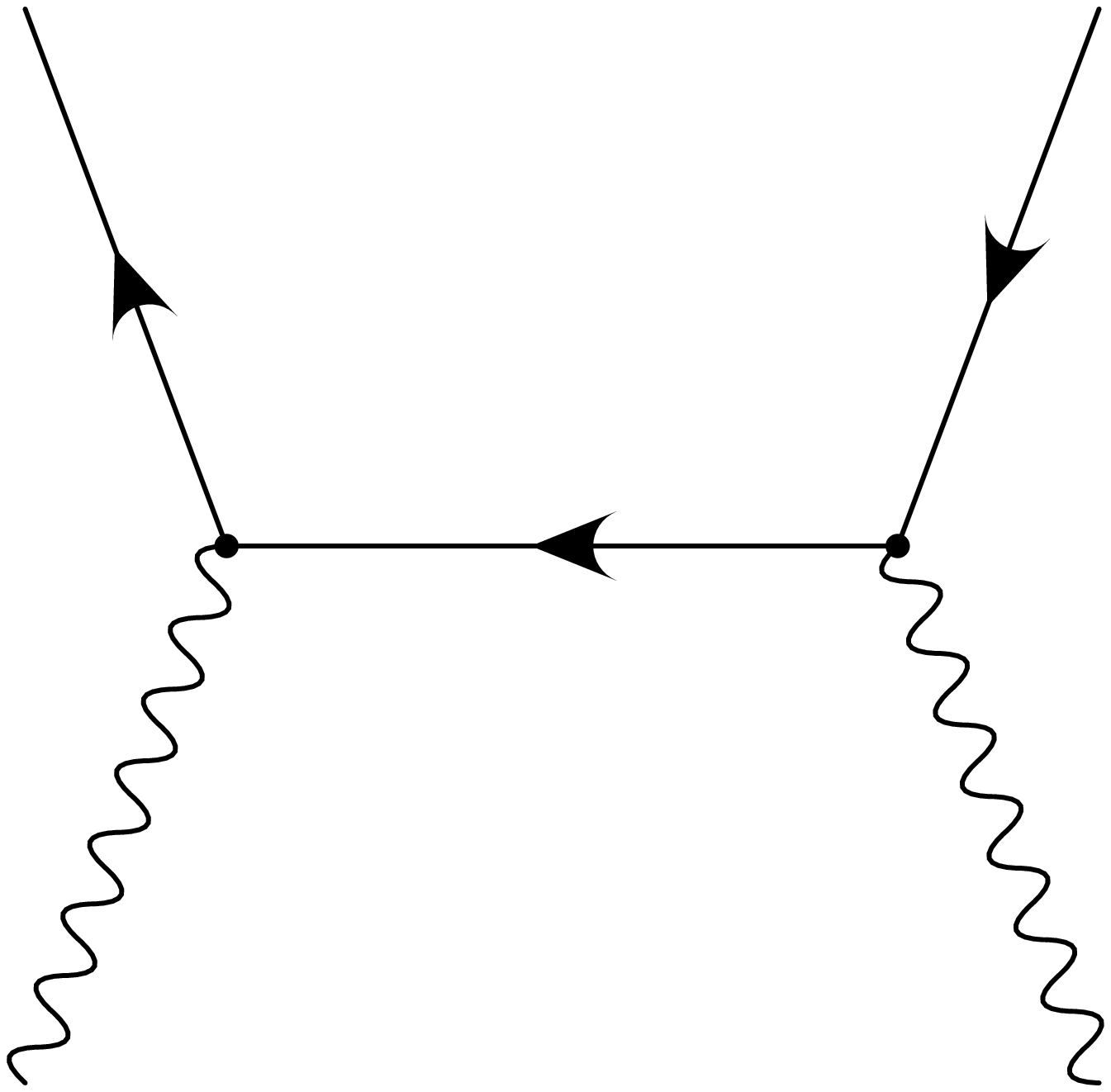}\quad
\includegraphics[width=2.5cm,angle=180]{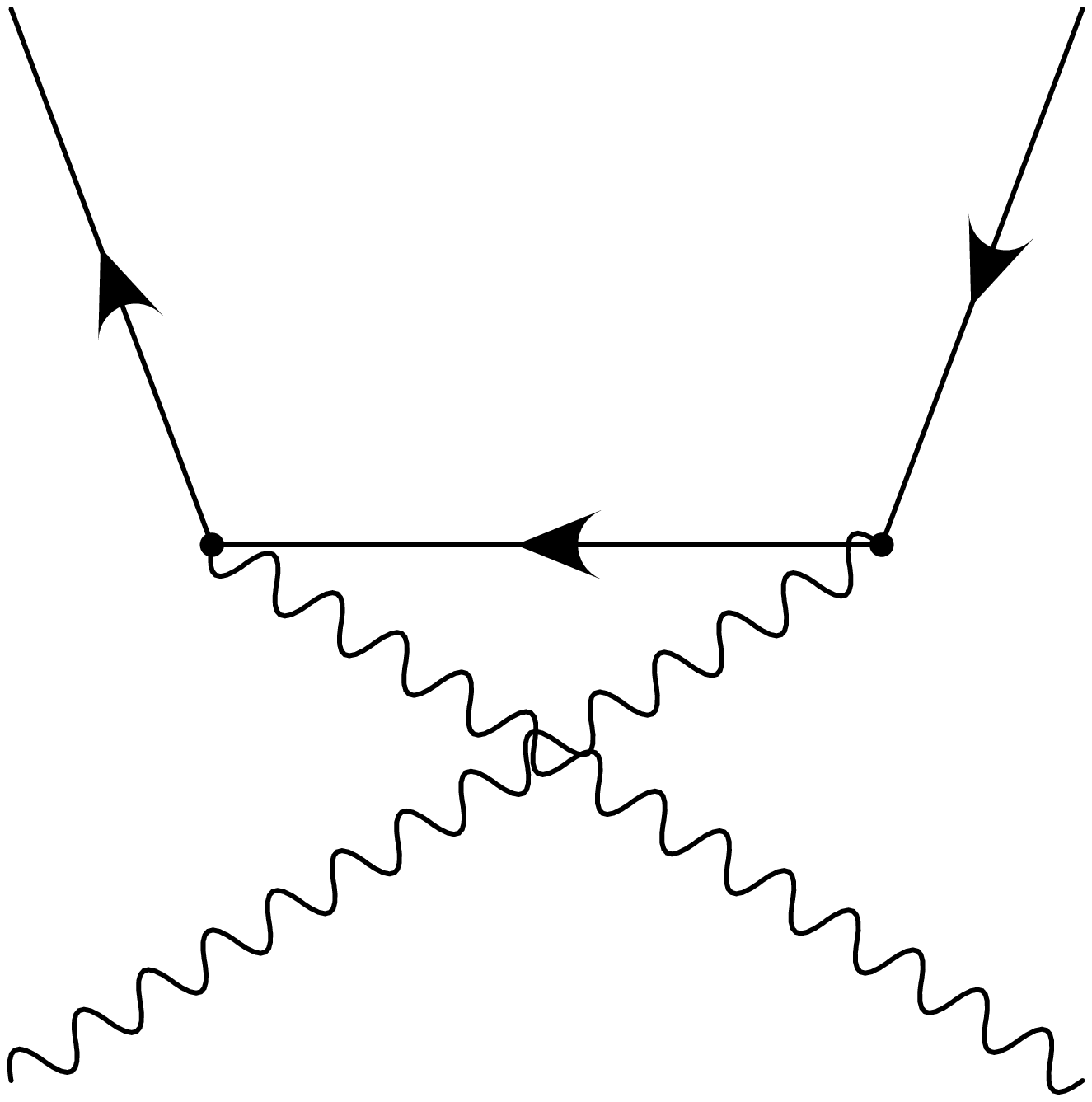}\quad
\includegraphics[width=2.5cm,angle=180]{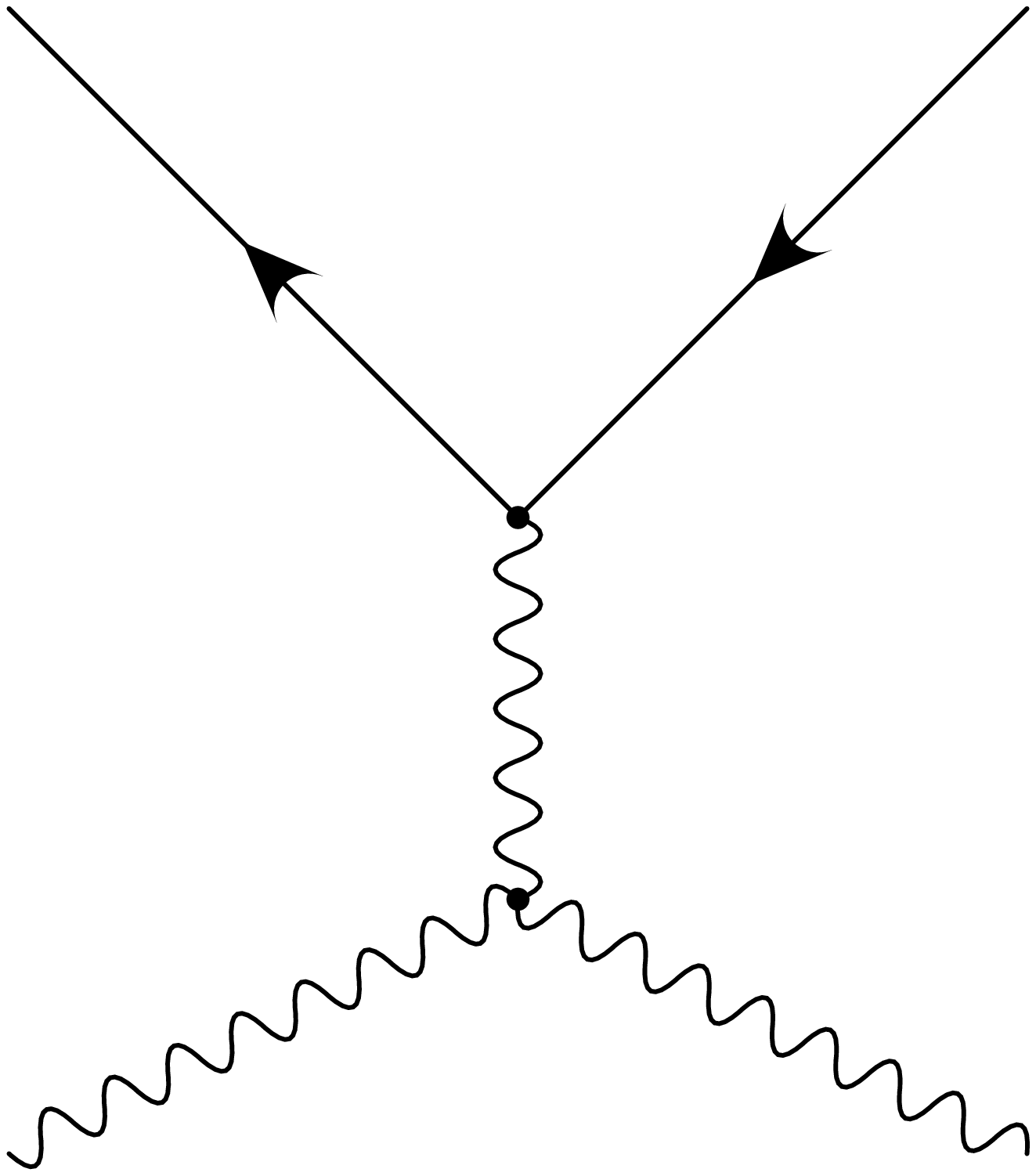}
\caption{ \label{fig:ggtt} Tree-level graphs for $gg \to q \bar q$. The first and second graphs have singularities for forward and backward
scattering, respectively.}
\end{figure}

In this section, we study the real and virtual corrections to heavy quark production via gluon fusion, $ g g \to q \bar q$. The tree-level graphs are given in Fig.~\ref{fig:ggtt}. The real radiation is computed by numerical integration using \mgv~\cite{Alwall:2014hca}. The virtual corrections use the SCET results of Ref.~\cite{Chiu:2009ft}. Since the real emission rate is a fixed order result, the virtual correction is expanded out to order $\alpha_W$ to study the real-virtual cancellation.

The $g g \to q \bar q$ total cross-section has a $t$-channel singularity for forward scattering, and a $u$-channel singularity for backward scattering, from the graphs in Fig.~\ref{fig:ggtt}. To avoid these singularities, we impose rapidity cuts. We require the particle with highest transverse momentum to have $\abs{\eta} < 1$ or $\abs{\eta} < 3$. We will refer to these as $\abs{\eta}<1,3$ cuts, respectively. We also require that the particle with second highest $p_T$ satisfy $\abs{\eta} < 5$. These cuts allow for collinear and soft $W$ emission from energetic quarks, but avoid the forward and backward singularities. They are applied to both the $gg \to q \bar q$ and $g g \to q \bar q W$ rates.

The scattering cross section can depend on the collision energy $s=E_{\text{CM}}^2$, the rapidity cut $\eta$, and the particle masses $\{M\}$. If the cross section is infrared finite as $\{M\} \to 0$, then it cannot contain $\ln s/M^2$ terms. The Sudakov logarithms are a sign that the cross section is divergent in the massless limit. In the $gg \to q \bar q$ case, the real and virtual corrections have Sudakov logarithms which cancel in the total rate.

We study the $gg \to q \bar q , q \bar q W $ rates for three cases:
\begin{enumerate}
\item $q=u,d$
\item $q=t,b$ with $m_b$=100\,GeV and $m_t=173$\,GeV
\item $q=t,b$ with $m_b$=4.7\,GeV and $m_t=173$\,GeV. 
 \end{enumerate}
Case (1) allows us to explain the structure of the gauge corrections without worrying about mass effects and Higgs corrections. Case (2) also involves Higgs radiative corrections, but has a stable $t$ quark since $m_t < m_b+m_W$. Finally case (3) is the physical case with an unstable $t$, which can decay via $ t \to b W$ decay.

The virtual corrections can be computed from the results in Ref.~\cite{Chiu:2009ft} (including also the $y_b$ terms), and are obtained by averaging the electroweak corrections for left- and right-handed quarks. The virtual corrections to the cross sections are
\begin{align}
\sigma_V(gg \to t \overline t)  &= \sigma_{0,t} \left\{  v_W + 3 v_t + v_b   \right\} \nn
\sigma_V(gg \to b \overline b)  &= \sigma_{0,b} \left\{ v_W + v_t + 3 v_b   \right\}
\label{eqns}
\end{align}
where
\begin{align}
v_W &=\frac{C_F \alpha_W}{4\pi} \left[ -\lL^2 + 3 \lL  \right] ,\nn
v_t &=- \frac{y_t^2}{32\pi^2} \lL, \nn
v_b  &=- \frac{y_b^2}{32\pi^2} \lL\,
\label{eqns1}
\end{align}
$\sigma_{0,t}=\sigma(gg \to t \bar t)$, and $\sigma_{0,b}=\sigma(gg \to b \bar b)$
are the corresponding tree-level rates,
$C_F=3/4$ for $SU(2)$, and $y_{t,b}$ are the quark Yukawa couplings. The corrections for $u,d$ quarks are given by $y_{t,b}\to0$. The tree-level  cross section $\sigma_0$ depends on the $\eta$ cut. The virtual rates depend on  the $\eta$ cut in the same way as the tree-level rates. The reason is that the virtual electroweak corrections for $g g \to q \bar q$ do not depend on the kinematic variables (such as the scattering angle) in this case, so the radiative correction is an overall multiplicative factor. In other cases, such as $q \overline q \to q \bar q$, the virtual electroweak corrections depend on kinematic variables, and have to be integrated over phase space. The gauge radiative corrections have both $\lL^2$ and $\lL$ terms, whereas the Higgs radiative corrections are linear in $\lL$.

\subsection{$u,d$ Quark Production}\label{sec:uu}

The tree-level processes are $gg \to u \bar u$ and $gg \to d \bar d$, and the real radiation processes are $gg \to u \bar u Z$, $gg \to d \bar d Z$, $gg \to u \bar d W^-$ and $gg \to d \bar u W^+$.   Since we are working in an $SU(2)_W$ theory (with $Z=W^3$), custodial $SU(2)$ implies that the $\sigma(u\bar u)=\sigma(d \bar d)$, and $\sigma(u \bar d W^-)=\sigma(d \bar u W^+)=2\sigma(u \bar u Z)=2\sigma(d \bar d Z)$.

\begin{figure}
{\includegraphics[width=\textwidth]{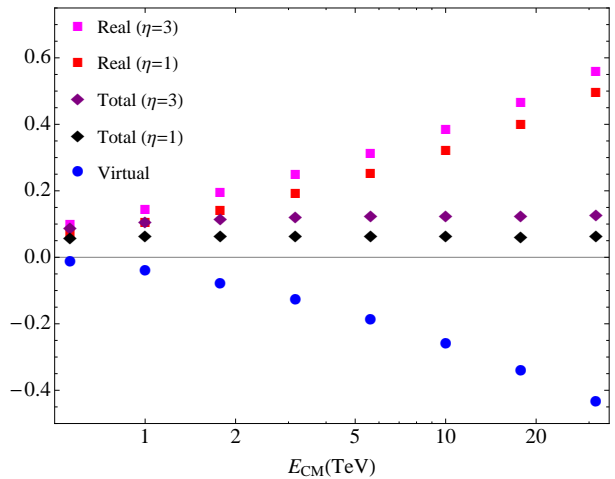}}
\caption{\label{fig:udw} Plot of real and virtual corrections to $gg \to q \bar q$ for $q=u,d$.  All rates have been normalized to the tree-level $gg \to u \bar u$ rate. The virtual correction to $g g \to q \bar q$ is shown as blue dots. The $gg \to q \bar q W$ real emission rate as a function of $E_{\text{CM}}$ for $\abs{\eta}<1,3$ cuts are shown as red and purple squares, respectively. The
$\alpha_W$ correction to the total rate with $\abs{\eta}<1$  and $\abs{\eta}<3$ cuts are shown as red and purple diamonds, respectively.
}
\end{figure}

Figure~\ref{fig:udw} shows the real and virtual corrections to the $u \bar u, d \bar d$ production rate, as a function of $E_{\text{CM}}$, for $\abs{\eta}<1,3$ cuts. All  rates have been normalized by dividing by the tree-level $gg \to u \bar u$ rate for the corresponding $\eta$ cut. This removes the overall $1/s$ dependence of the cross sections. The graph clearly shows that the virtual and real cross sections become large at high energy, and the $\lL^2$ dependence is reflected in the quadratic shape of the curves.  The virtual correction is independent of the $\eta$ cut, and as is typical of Sudakov effects, is negative. The real correction depends on the $\eta$ cut. The $\lL^2,\lL$ corrections arise from soft and collinear radiation; the real radiation kinematics for the final state quarks in $gg \to q\bar q W$ is similar to that for the tree-level $gg \to q \bar q$ process.
As a result, the $\lL^2,\lL$ terms do not depend on the $\eta$ cut, and only the constant $\lL^0$ term does. This is reflected in the figure by the fact that the difference in cross sections between the two values of the $\eta$ cut remains constant as $E_{\rm CM}$ is changed. 

The $\lL^2,\lL$ terms cancel in the total cross section, as is evident by the curves for the total rate becoming horizontal for large energy, and only the constant terms survive. The electroweak corrections to the total cross section are at the 10\% level. At  partonic center-of-mass energies of about one TeV, the individual corrections from the real and virtual corrections are also at the 10\% level, but they rise quickly as $E_{\rm CM}$ is increased. 

For a 100~TeV machine, partonic center-of-mass energies can exceed 10\,TeV, and the corrections become large (factors of 2). For most experimentally relevant processes there is never a complete cancellation of the logarithms (since one is typically not measuring a totally inclusive rate, and furthermore the initial state is not an $SU(2)$ singlet), the resummed expressions are needed. 

The cancellation between real and virtual corrections is
\begin{align}
3 \sigma(u \bar d W) +2  v_W  \sigma(u \bar u) & \to 0
\label{c1}
\end{align}
using the isospin relations mentioned earlier and Eqs.~(\ref{eqns},\ref{eqns1}),  where $\to 0$ means that the $\lL^2,\lL$ dependence cancels, but there can be constant terms left over.

It is important to note that for initial states that are not electroweak singlets, such as for $q \bar q \to q \bar q$, the real and virtual corrections  have \emph{different} $\lL^2,\lL$ dependence, and the large corrections persist in the total cross section. This non-cancellation persists even at the hadron level. The $pp \to t \bar t$ rate has large corrections from the $q  \bar q \to q \bar q$ channel, since the $u$ and $d$ quark distributions in the proton are not the same.

\subsection{$t,b$ Quark Production with $m_b=100$\,GeV}\label{sec:100}

\begin{figure}
{\includegraphics[width=\textwidth]{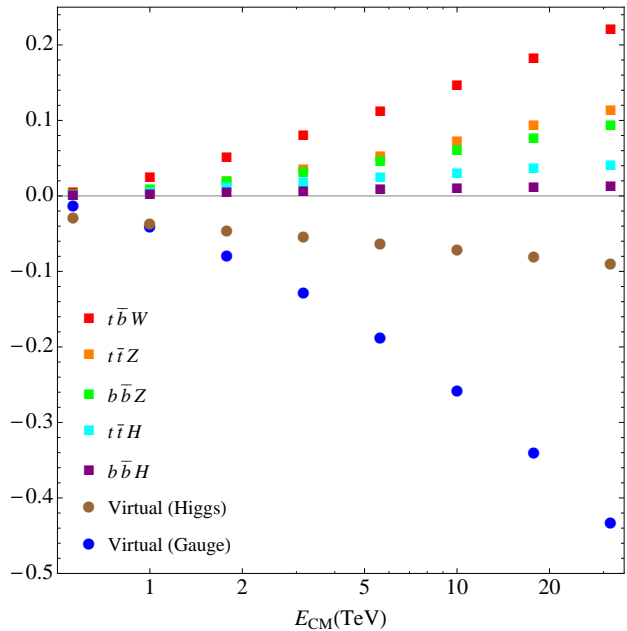}}
\caption{Plot of real and virtual corrections to $gg \to q \bar q$ for $q=t,b$ for $m_b=100$\,GeV with an $\abs{\eta}<1$ cut. All rates have been normalized to the tree-level $gg \to u \bar u$ rate. The points are: virtual correction gauge corrections (blue dots), virtual Higgs corrections (brown dots), $t \bar b W^-$ (red squares), $t \bar t Z$ (orange squares), $b \bar  b Z$ (green squares), $t \bar t H$ (cyan squares) and $b \bar b H$ (purple squares).
\label{fig:tbw100c}}
\end{figure}

\begin{figure}
{\includegraphics[width=\textwidth]{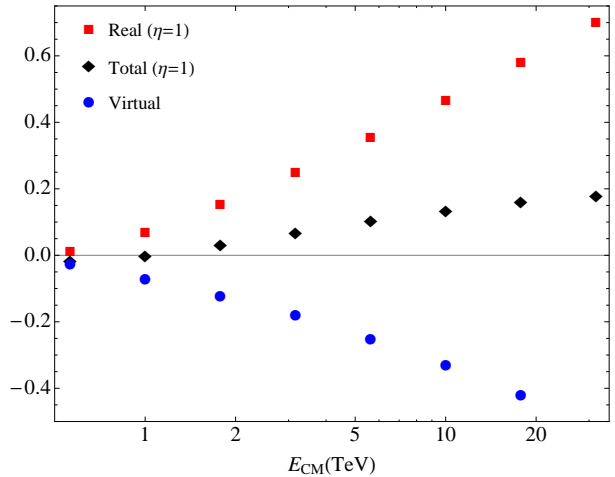}}
\caption{Plot of the real and virtual corrections to $gg \to q \bar q$ for $q=t,b$ for $m_b=100$\,GeV with an $\abs{\eta}<1$ cut. The virtual corrections are blue dots, the total real emission rate is shown as red squares, and the total radiative correction is shown as black diamonds. All rates are normalized to the $gg \to u \bar u$ rate. The total rate levels off beyond 30\,TeV.
\label{fig:tbw100a}}
\end{figure}

We now consider the case of $gg \to t \bar t, b \bar b$ for $m_t=173$\,GeV and $m_b=100$\, GeV. An unphysical $b$ mass has been chosen, so that the $t \to b W$ decay is forbidden.  The case of unstable top is discussed in Sec.~\ref{sec:47}. The virtual corrections for $t \bar t$ and $b \bar b$ production are given in Eq.~(\ref{eqns}). The real rates are computed using \mgv. All rates are divided by the corresponding $gg \to u \bar u$ rate to remove an overall $1/s$ normalization factor. The tree-level rates $gg \to t \bar t$ and $gg \to b \bar b$ are essentially equal to $gg \to u \bar u$ except very close to $t \bar t$ threshold, so each of these tree-level rates are $1$ in the normalization of the plot, and have not been shown.

The real and virtual corrections are shown in Fig.~\ref{fig:tbw100c} for the $\abs{\eta}<1$ cut. The $\abs{\eta}<3$ plots are very similar, with a small offset from the $\abs{\eta}<1$ curves, as for the $u,d$ case in Fig.~\ref{fig:udw}. The $t \bar b W^-$ emission rate is the sum of the rates for transversely and longitudinally polarized gauge bosons. The rate for transversely polarized gauge bosons at high energies is the same as that for $u \bar d W^-$ production, since fermion mass effects are power suppressed. The rate for longitudinally polarized gauge bosons is the same as for emission of the unphysical scalar (by the equivalence theorem), and is related to the Higgs emission rate. The real and virtual rates can be written in terms of the $u\bar d W^-$ rate and the rate $\sigma_S$ to emit a scalar with unit Yukawa coupling,
\begin{align}
\sigma(t \bar b W^-)  &\to \sigma(u \bar d W^-) + 2 (y_t^2+y_b^2) \sigma_S \nn
\sigma(t \bar t Z)  &\to \frac12 \sigma(u \bar d W^-) + 2 y_t^2 \sigma_S \nn
\sigma(b \bar b Z)  &\to \frac12 \sigma(u \bar d W^-) + 2 y_b^2 \sigma_S \nn
\sigma(t \bar t H)  &\to 2 y_t^2 \sigma_S \nn
\sigma(b \bar b H)  &\to 2 y_b^2 \sigma_S \nn
\sigma_V(t \bar t) &\to \left(v_W + 3 v_t + v_b\right)\sigma(u \bar u) \nn
\sigma_V(b \bar b) &\to \left(v_W +  v_t + 3 v_b\right)\sigma(u \bar u) 
\label{22}
\end{align}
The $\sigma(u \bar d W^-)$ terms in $\sigma(t \bar b W^-)$, etc., are for transverse $W$ and $Z$ emission and the $\sigma_S$ terms are for longitudinal $W$ and $Z$ emission.\footnote{Remember that $Z=W^3$ since we are in a pure $SU(2)_W$ theory. Otherwise, the $Z$ rates would have additional factors of $1/\cos^2 \theta_W$.} One can verify that the real emission curves in Fig.~\ref{fig:tbw100c} satisfy Eq.~(\ref{22}), so that five curves are given in terms of two quantities, $\sigma(u \bar d W^-)$ determined already in Sec.~\ref{sec:uu}, and $\sigma_S$. The Higgs emission curves $\sigma(t \bar t H),\sigma(b \bar b H) $ are linear, which means  they contain $\lL$ terms but no $\lL^2$ terms.

The sum of all the real radiation rates, as well as the total cross section, are shown in Fig.~\ref{fig:tbw100a}. The total cross section levels out at high energy (we have verified  this by continuing the plot to even higher center of mass energies), which shows numerically that the $\lL^2$ and $\lL$ terms cancel between the real and virtual corrections. The total real emission rate is
\begin{align}
 \sigma_R &=2 \sigma(t \bar b W^-)+ \sigma(t \bar t Z) + \sigma(b \bar b Z) + \sigma(t \bar t H)  +\sigma(b \bar b H)  \nn
 &\to 3 \sigma(u \bar d W^-) + 8 (y_t^2+y_b^2) \sigma_S 
\end{align}
and the total virtual rate is
\begin{align}
 \sigma_V &=\sigma_V(t \bar t)+\sigma_V(b \bar b)   = \left(2 v_W + 4 v_t + 4 v_b\right)\sigma(u \bar u) 
\end{align}
The cancellation $\sigma_R +\sigma_V \to 0$ implies that
\begin{align}
3 \sigma(u \bar d W^-) + 8 (y_t^2+y_b^2) \sigma_S  + \left(2 v_W + 4 v_t + 4 v_b\right)\sigma(u \bar u)  \to 0.
\end{align}
The gauge and Higgs parts cancel separately. The gauge part cancels using Eq.~(\ref{c1}), and
\begin{align}
8 (y_t^2+y_b^2) \sigma_S  + \left(4 v_t + 4 v_b\right)\sigma(u \bar u)  \to 0.
\label{c2}
\end{align}
From Eq.~(\ref{eqns1}), we see that $v_{t,b}$ are linear in $\lL$, which explains the linearity of the Higgs emission cross section $\sigma_S$. 

\subsection{$t,b$ Quark Production with $m_b=4.7$\,GeV}\label{sec:47}

\begin{figure}
{\includegraphics[width=\textwidth]{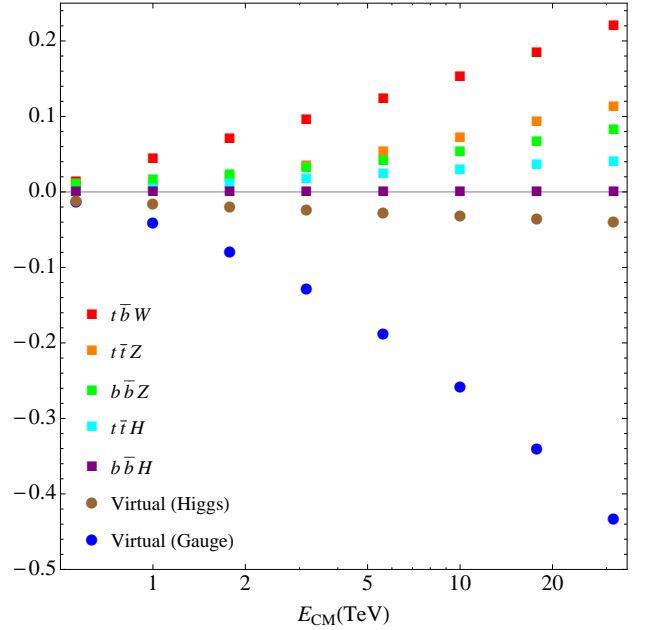}}
\caption{Same as Fig.~\ref{fig:tbw100c}, but for $m_b=4.7$\,GeV.
\label{fig:tbw47c}}
\end{figure}

\begin{figure}
{\includegraphics[width=\textwidth]{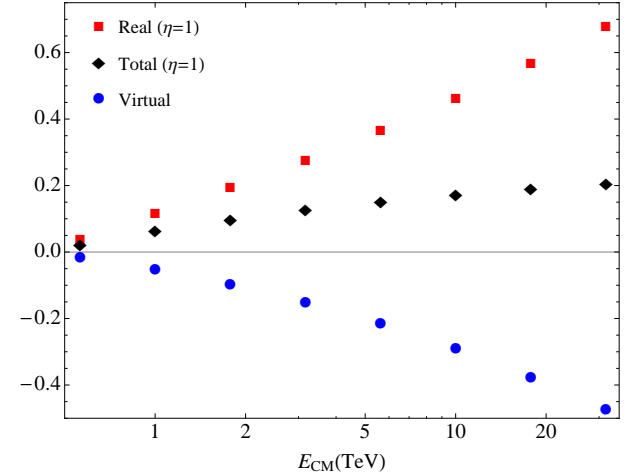}}
\caption{Same as Fig.~\ref{fig:tbw100a}, but for $m_b=4.7$\,GeV.
\label{fig:tbw47a}}
\end{figure}

\begin{figure*}
\includegraphics[bb=232 365 380 405,width=1.8cm]{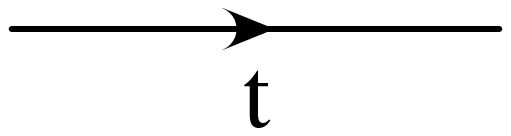}
\raise0.3cm\hbox{$\ +\ $}
\includegraphics[bb=178 365 434 509,width=3cm]{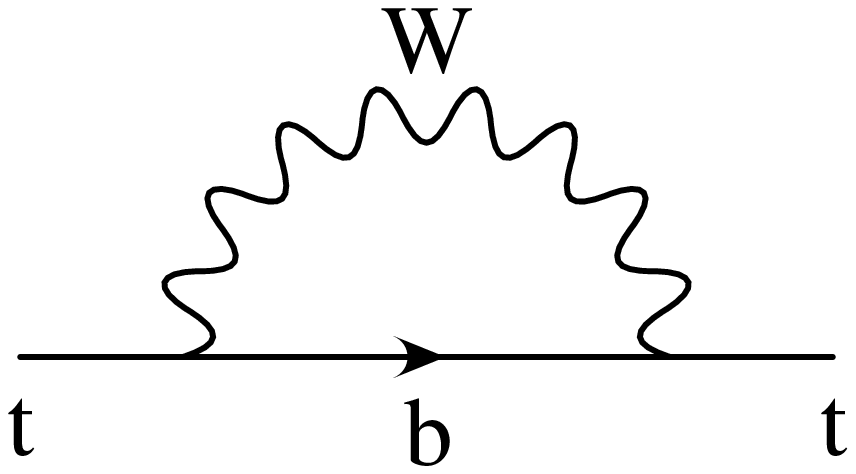}
\raise0.3cm\hbox{$\ +\ $}
\includegraphics[bb= 68 365 544 480,width=6cm]{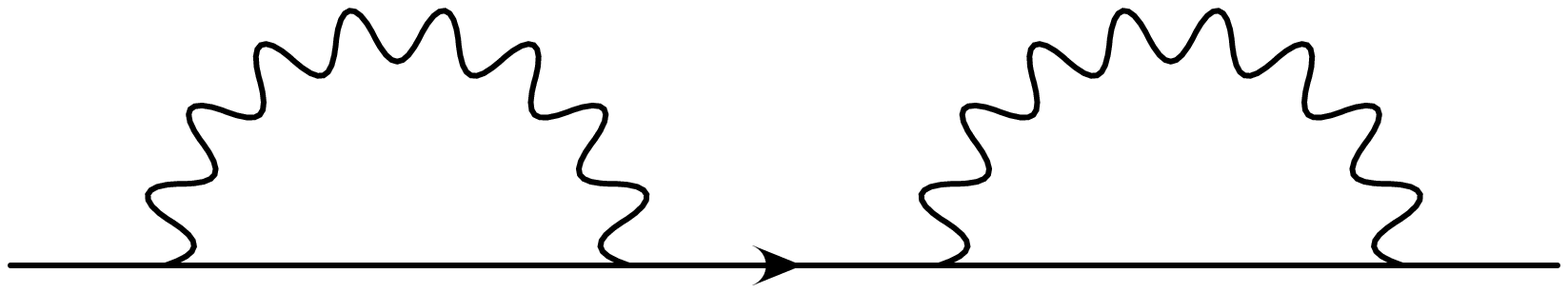}
\raise0.3cm\hbox{$\ +\ \ldots$}

\caption{\label{fig:nwa} Graphs that are summed in the narrow width approximation. In Eq.~(\ref{width}), only the imaginary part
of each loop is included.}
\end{figure*}

Finally, we study the case of a physical $b$ quark with $m_b=4.7$\,GeV and an unstable $t$ quark. The virtual corrections are still given by Eq.~(\ref{eqns}). There is, however, an important change in the $t \bar b W^-$ decay rate because the process $g g \to t \bar t$ followed by $\bar t \to \bar b W^-$ contributes to this rate. The $t \bar b W^-$ differential decay rate has a singularity when $(p_{\bar b}+p_{W^-})^2=m_t^2$, and the cross section diverges when integrated over final state phase space. The standard way to resolve this singularity is to regulate it by the $t$-quark width  using the replacement (the narrow width approximation, which is what is used in \mgv)
\begin{align}
\frac{1}{p^2-m_t^2+i\epsilon} \to  \frac{1}{p^2-m_t^2+i m_t \Gamma_t} 
\label{width}
\end{align}
for the $t$-quark propagator, where $\Gamma_t$ is the $t$-quark width. This is equivalent to summing a class of diagrams, the imaginary parts of $W$ corrections to the $t$-quark propagator, shown in Fig.~\ref{fig:nwa}. This is not gauge invariant, and also formally mixes different orders in the $\alpha_W$ expansion, since the $t$-quark width is $\mathcal{O}(\alpha_W m_t)$. The cut in the second graph of Fig.~\ref{fig:cut} is the same cut as occurs in summing the imaginary parts of Fig.~\ref{fig:nwa}, and the two cuts cannot be treated separately, as is done in the narrow width approximation.

If the $t \to b W^-$  decay is kinematically forbidden, the  $ t \overline b W^-$ real emission rate is order $\alpha_W$. When the decay is kinematically allowed, the $t \overline b W^-$ rate becomes order $1$. The reason is that in the resonance region, the rate is enhanced by a factor of $1/\Gamma_t$. The total  $t \overline b W^-$ rate includes what, in the kinematically forbidden case, is the $\mathcal{O}(1)$ $t \bar t$ rate. Once the $ t \overline b W^-$  decay is kinematically allowed, the approximation Eq.~(\ref{width}), while getting the correct $\mathcal{O}(1)$ rate, does not get the correct $\mathcal{O}(\alpha_W)$ piece.

To understand how the infrared divergence cancellation occurs for an unstable $t$ quark, consider the simpler case of $t \bar t$ production by a current $J$, as in Sec.~\ref{sec:cancel}. The $\alpha_W$ correction to the total rate can be computed from the imaginary part of the vacuum polarization graphs in Fig.~\ref{fig:cut}. The vacuum polarization $\Pi(q^2)$ has no  singularities for Euclidean $q^2$ even if $m_t > m_b + m_W$, so the analytic continuation to timelike $q^2$ does not either. The imaginary part for timelike $q^2$ is given by the real emission and virtual correction cuts shown in Fig.~\ref{fig:cut}, so the two contributions combined have no infrared divergence.

The graphs in Fig.~\ref{fig:cut} are all order $\alpha_W$, and their total gives the $\mathcal{O}(\alpha_W)$  correction to the total rate. The  graphs are computed with the $t$-quark propagator on the l.h.s. of Eq.~(\ref{width}), rather than the narrow width approximation on the r.h.s. The real emission graph is singular because the $ t \to  b W^-$ decay is kinematically allowed. A careful calculation shows that the virtual correction is also singular, and the sum is finite. The cancellation can be checked using the l.h.s.\ of Eq.~(\ref{width}) with the $i\epsilon$ term acting as a regulator. The real and virtual graphs each have a piece proportional to $1/\epsilon$, which cancels in the sum.

\begin{figure}
\includegraphics[width=0.9\textwidth]{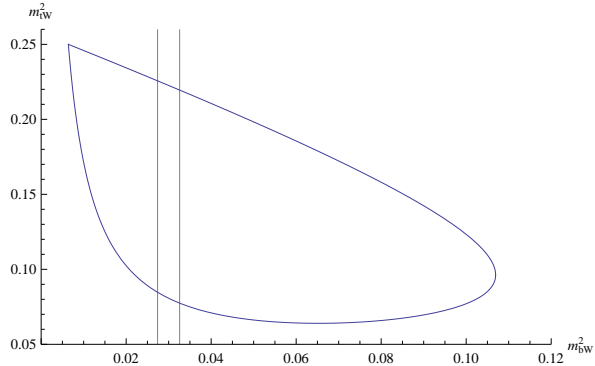}
\caption{\label{fig:ps} Phase space region for $t \bar b W$ production for $E_{\text{CM}}=500$\,GeV. The vertical band is the region where $(m_t-5\Gamma_t)^2 \le m^2_{bW} \le (m_t+5\Gamma_t)^2 $. The axes are in $(\text{TeV} )^2$.}
\end{figure}

The $t \bar b W^-$ rate can be computed by adding the rates for two regions: $A$, which is a small region around where the $t$-quark is on-shell, and $A^\prime$, which is the rest of phase space. In terms of the final state phase space variables $m^2_{bW}=(p_b+p_W)^2$, $m^2_{tW}=(p_t+p_W)^2$ needed for three-body decay,  $A$ is the region $m_t^2-\Delta \le m^2_{bW} < m_t^2+\Delta$, and $A^\prime$ is the remaining region. The phase space region is shown in Fig.~\ref{fig:ps},  with $A$ the region within the vertical band, and $A^\prime$ outside. For a stable $t$-quark, the vertical band moves outside the allowed phase space region, and there is no singularity in the phase space integral. For an unstable $t$-quark, the rate is non-singular in region $A^\prime$, and can be computed by the propagator on the l.h.s.\ of Eq.~(\ref{width}). To correctly compute the $\mathcal{O}(\alpha_W)$ terms, one must also use the propagator on the l.h.s.\ of Eq.~(\ref{width}), rather than the narrow width approximation on the r.h.s., for the integral over the singular region $A$. 

The region $A$ contribution has a singular $1/\epsilon$ piece that must be subtracted, keeping only the finite $\mathcal{O}(\alpha_W)$ part. The  $1/\epsilon$ singular part of the rate becomes the $\mathcal{O}(1)$ contribution in the narrow width approximation, and the subleading  $\mathcal{O}(\alpha_W)$ is, unfortunately, not given correctly by the narrow width approximation.

The phase space integral of the decay distribution over $A$ has the form
\begin{align}
I &=\int_{m_t^2-\Delta}^{m_t^2+\Delta}  \rd m^2_{bW}\ \frac{ f(m^2_{bW})}{(m^2_{bW}-m_t^2)^2+\epsilon^2}
\end{align}
where the denominator is from the absolute value squared of the propagator in Eq.~(\ref{width}),  $f$ contains all  non-singular factors in the decay distribution, and $\Delta$ is the width of the integration region. Expanding around $m_t^2$,
\begin{align}
 f(m^2_{bW})=f_0 +  (m^2_{bW}-m_t^2) f_1 +(m^2_{bW}-m_t^2)^2 f_2  + \ldots
\end{align}
gives
\begin{align}
I &= \frac{\pi}{\epsilon} f_0 + 2 \Delta f_2 + \ldots
\end{align}
The first term is the singular $1/\epsilon$ piece that must be subtracted, and the remaining terms are the finite $\mathcal{O}(\alpha_W)$ terms. Relative to the contribution from region $A^\prime$, they are smaller by a factor $\Delta$, i.e.\ the width of the vertical band relative to the width of the full phase space region. Since we are only interested in the $\mathcal{O}(\alpha_W)$ contribution to the rate, we can get a good estimate of this by simply using the contribution from region $A^\prime$, and ignoring $A$. The $\mathcal{O}(\alpha_W)$ term from $A$ is a small correction, since the size of $A$ is much smaller than $A^\prime$. A practical way to do this in \mgv\ is to use the {\tt \$t} tag, which excludes a region of width $15\Gamma_t$ around the on-shell $t$-quark.

The results of this computation are shown in Figs.~(\ref{fig:tbw47c},\ref{fig:tbw47a}), and are very similar to those for $m_b=100$\,GeV. The main difference is the Yukawa correction is smaller, since $y_b$ is now almost zero. The entire discussion of Sec.~\ref{sec:100} holds, and will not be repeated again.

\section{Discussion and Conclusions}\label{sec:conclusions}

We have presented the electroweak radiative corrections to $gg \to t \bar t, gg  \to b \bar b$ production in Sec.~\ref{sec:tt}. The individual processes that contribute have large electroweak corrections that depend on $\lL^2$ and $\lL$, but these cancel in the total rate. The virtual corrections are around $-10$\% for $E_{\text{CM}} \sim 2$\,TeV, and grow with energy. 

The electroweak corrections to the individual processes are relevant for measurements at the LHC. For example, suppose one is interested in measuring the $t \bar t$ production rate. The virtual corrections to $t \bar t$ contribute to this rate. If one has a perfect detector, then one can exclude the real emission final states $t \bar b W$, $t \bar t Z$, $b \bar b Z$, $t \bar t H$, $b \bar b H$. In this case, the cross section is given by the blue dots in Fig.~\ref{fig:tbw47c}, and there are large electroweak radiative corrections. In a more realistic case, there will be some leakage from the real radiation processes into the $t \bar t$ channel. For example $t \bar t Z$ with $Z \to \nu \bar \nu$ could be mistaken for $t \bar t$, or $t \bar t Z$ with $Z \to q \bar q$, where the $Z$ decay products cannot be separated from the $t$-quark decay jets. If some fraction of the real radiation is included, then there will be some cancellation with the electroweak corrections to the virtual rate, so that the overall 
electroweak correction is somewhat smaller. A realistic calculation of the measured rates is beyond the scope of this work. To do such a calculation requires taking the corrections discussed in this paper, integrating over the gluon PDFs, and then putting the parton processes through a showering algorithm and detector cuts. In addition, one should also include the quark production rates $q \bar q \to t \bar t$, which were included in the analysis of Ref.~\cite{Manohar:2012rs}. As noted earlier, the electroweak corrections to $q \bar q \to t \bar t$ do not cancel even for the totally inclusive rate. It should be clear that even in a complete calculation, the electroweak corrections do not cancel, and a significant correction remains.

The electroweak radiative corrections start to become measurable at LHC energies, and their importance grows with energy. We have numerically studied the $g  g \to t \bar t$ process in this paper. Most processes have much larger electroweak corrections than this process, because they typically contain more particles with electroweak interactions. (The gluon does not have electroweak interactions at leading order.)
 The corrections for $q \bar q \to t \bar t$ are approximately twice as large, because the initial and final states both have electroweak interactions. Processes such as $q \bar q \to WW$ which involve electroweak gauge bosons have even larger corrections, since the group theory factor $C_F=3/4$ is replaced by $C_A=2$ in the amplitude.

The effective theory method breaks the electroweak correction into the high-scale matching $C$, the running $\gamma$ and the low-scale matching $D$. The $\lL^2$ term arise from $\gamma$, and the $\lL$ terms from $\gamma$ and $D$. All terms are known to NLL order, as are the most important terms at NNLL order (see appendix). 

In addition to the electroweak corrections, there are of course, QCD corrections, which are much larger, and have been included in existing calculations and implemented in Monte Carlo code. The QCD and electroweak corrections factor in $A$ and $D_L$ to two-loop order and in $B$, $D_0$ and $C$ to one-loop order~\cite{Chiu:2008vv,Chiu:2009mg}, so that the total radiative correction to NLL order can be written as the product $R_{\text{QCD}} R_{\text{EW}}$ . $R_{\text{QCD}}$ has been included in existing calculations, so the electroweak corrections can be included to NLL order simply by reweighing the QCD results by $R_{\text{EW}}$. This has to be done before integrating over the final state phase space, since $R_{\text{EW}}$ can depend on kinematic variables such as scattering angles. One complication is that $R_{\text{EW}}$ depends on the helicities of the partons, since the weak interactions are chiral.

The experimental energy reach at the LHC is high enough that electroweak corrections should be included in measurements that are approaching 10\% accuracy. Recently, there have been studies of a possible 100\,TeV hadron collider. At these high energies, the electroweak corrections are large, and must be resummed to have reliable cross sections.

\begin{acknowledgements}

 BS would like to thank Rodrigo Alonso for comments on the manuscript. BS and AM were supported in part by DOE grant DE-SC0009919. 
 The research of ST was supported by the Lawrence Berkeley National Laboratory and he thanks them for their hospitality, by a DFG Forschungsstipendium under contract no.~TU350/1-1 and by
 ERC Advanced Grant EFT4LHC of the European Research Council and the Cluster
of Excellence Precision Physics, Fundamental Interactions and Structure of
Matter (PRISMA-EXC 1098).

\end{acknowledgements}

\begin{appendix}

\section{Summary of \scetew\ Results}\label{sec:scet}

We now summarize the results of Refs.~\cite{Chiu:2007yn,Chiu:2007dg,Chiu:2008vv,Chiu:2009yz,Chiu:2009yx,Chiu:2009mg,Chiu:2009ft,Fuhrer:2010eu} for the electroweak corrections.

\begin{enumerate}

\item At a high scale $\mu_h$ of order $s$, the scattering amplitudes are matched onto $SU(3) \times SU(2) \times U(1)$ gauge invariant local operators $O_i$ with coefficients $C_i$ which can be computed perturbatively in a power series in $\alpha(\mu_h)$. The calculations in Refs.~\cite{Chiu:2007yn,Chiu:2007dg,Chiu:2008vv,Chiu:2009yz,Chiu:2009yx,Chiu:2009mg,Chiu:2009ft,Fuhrer:2010eu}  include QCD as well as electroweak corrections, so $\alpha$ denotes any of the three gauge coupling constants in the Standard Model (SM). As an example, for $g(p_1)+g(p_2) \to q(p_3)+\overline q(p_4)$, the operators are
\begin{align}
O_1 &=  \bar q_4 q_3 A_2^A A_1^A\nn
O_2 &=  d^{ABC} \bar q_4 T^C q_3 A_2^A A_1^B\nn
O_3 &= if^{ABC} \bar q_4 T^C q_3 A_2^A A_1^B\,.
\label{ops}
\end{align}
which give the possible color structures of the amplitude. The subscripts $1,2,3,4$ label the different particle momenta.

\item The coefficients $C_i$ are evolved using renormalization group equations (RGE) down to a low scale $\mu_l$ of order $M_W$. The anomalous dimensions can be computed in the unbroken $SU(3) \times SU(2) \times U(1)$ theory.

\item At the scale $\mu_l$, the $W$, $Z$, $H$ and $t$ are integrated out. This calculation must be done in the broken theory.
 A single gauge invariant operator breaks up into different components because the weak interaction symmetry is broken. For example, each of the operators $O_i$ in Eq.~(\ref{ops}) breaks up into an $SU(3)$ invariant $gg \to t \overline t$ and $gg \to b \overline b$ operator.
 
\item The operators in the theory below $\mu_l$ are then used to compute the scattering cross sections. 

\end{enumerate}

The final result is that the scattering amplitudes $\mathcal{M}$ can be written as
\begin{align}
\mathcal{M} &=  \exp \left[D_C(\mu_l,\lM,\bar n \cdot p) \right] d_S(\mu_l,\lM)\nn
&\times P\,\exp \left[\int_{\mu_h}^{\mu_l} \frac{\rd \mu}{\mu} \gamma(\mu,\bar n \cdot p) \right] C(\mu_h,\lQ)
\label{m}
\end{align}
Eq.~(\ref{m}) gives the scattering amplitude in resummed form. Explicit formul\ae\ for all the pieces can be found in Ref.~\cite{Chiu:2009ft}.

The high-scale matching $C(\mu_h,\lQ)$ is an $n$ dimensional column vector with a perturbative expansion in $\alpha_i(\mu_h)$, with $i=1,2,3$ being the $U(1)$, $SU(2)$ and $SU(3)$ couplings. It also depends on $\lQ=\ln s/\mu_h^2$, which is not a large logarithm if one picks $\mu_h^2 \sim s$. For Eq.~(\ref{ops}), $n=3$ since there are 3 gauge invariant amplitudes. 

The SCET anomalous dimension $\gamma(\mu)$ is an $n \times n$ anomalous dimension matrix which can be written as the sum of a collinear and soft part
\begin{align}
\gamma(\mu,\bar n \cdot p) &= \gamma_C(\mu,\bar n \cdot p)+ \gamma_S(\mu)
\end{align}
where the collinear part is diagonal
\begin{align}
\gamma_C(\mu,\bar n \cdot p)&= \openone \sum_r \left[A_r(\mu) \ln \frac{2E_r}{\mu}+B_r(\mu) \right] 
\end{align}
 and linear in $\log \bar n_r \cdot p_r=E_r$, the energy of the parton, to all orders in perturbation theory~\cite{Manohar:2003vb,Chiu:2009mg}. The sum on $r$ is over all partons in the scattering process, and $A_r(\mu)$ and $B_r(\mu)$ have a perturbative expansion in $\alpha_i(\mu)$. $\gamma_S$ at one-loop order is
\begin{align}
\gamma_S(\mu) &= -\sum_{\vev{rs},i} \frac{\alpha_i(\mu)}{\pi} T_r^{(i)} \cdot T_s^{(i)}  \ln \frac{-n_r \cdot n_s + i 0^+}{2}
\end{align}
where the sum is over all parton pairs $\vev{rs}$, and $n_r=(1,\mathbf{n}_r)$ is a null vector in the direction of parton $r$ for each incoming parton, and $n_r=-(1,\mathbf{n}_r)$ for each outgoing parton. $T_r^{(i)}$ is the gauge generator for the $\imath^{\rm th}$ gauge group acting on parton $r$.

The low-scale matching  has a collinear part $D_C$ and a soft part $d_S$. The soft part $d_S$ is an $m \times n$ matrix, where $m$ is the number of amplitudes produced after $SU(2) \times U(1)$ breaking. In $gg \to q \overline q$, if $q$ is an electroweak doublet of left-handed quarks $(t,b)_L$, then starting with the operators in Eq.~(\ref{ops}) gives $m=6$ operators after $SU(2) \times U(1)$ breaking, where $\overline q_4 q_3 \to \overline t_4 t_3$, or $\overline q_4 q_3 \to \overline b_4 b_3$. If $q$ in Eq.~(\ref{ops}) is an electroweak singlet, such as $b_R$ or $t_R$, then $m=3$. $d_S(\mu,\lM)$ has an expansion in $\alpha_{S,W,\text{EM}}(\mu_l)$, and can depend on electroweak scale masses and $\mu_l$ via dimensionless ratios such as $M_W/M_Z$ and $\lM=\ln M_Z/\mu_l$. The logarithms are small if one chooses $\mu_l \sim M_Z$.

The collinear matching $D_C$ is an $m \times m$ diagonal matrix given by
\begin{align}
\left[D_C(\mu,\bar n \cdot p,\lM)\right]_{ii} &= \sum_r \left[ J_r (\mu,\lM)\ln \frac{2E_r}{\mu} + H_r (\mu,\lM) \right]
\label{DC}
\end{align}
and $J_r$ and $H_r$ are functions of $\alpha_{S,W,\text{EM}}(\mu_l)$, and can depend on electroweak scale masses and $\mu_l$ via dimensionless ratios such as $M_W/M_Z$ and $\lM=\ln M_Z/\mu_l$.  The sum on $r$ is over all particles in operator $O_i$ produced after electroweak symmetry breaking, and $D_C$ is linear in $\ln \bar n \cdot p$ to all orders in perturbation theory~\cite{Manohar:2003vb,Chiu:2009mg}.

The exponent contains at most a double-log given by integrating the $A_i$ terms in the collinear anomalous dimension. The low-scale matching contains a single-log term. This a new feature of \scetew\ first pointed out in Ref.~\cite{Chiu:2007yn}. One can show that the low-scale matching contains at most a single-log to all orders in perturbation theory~\cite{Chiu:2007yn,Chiu:2009ft}. As a consequence, resummed perturbation theory remains valid even at high energy, because $\alpha^n \ln s/M_W^2 \ll 1$ for large enough $n$. $A_i$, $\gamma_S$, and $J_i$ are related to the cusp anomalous dimension.

The $\log$ term in the matching  Eq.~(\ref{DC}) is needed for proper factorization of scales. A typical Sudakov double-log term at one loop has the form (dropping the overall $\alpha$)
\begin{align}
 \ln^2 \frac{Q^2}{M^2} &=\ln^2 \frac{Q^2}{\mu_h^2} + \left[\ln^2 \frac{Q^2}{\mu_l^2}-\ln^2 \frac{Q^2}{\mu_h^2} \right]\nn
 &+\left[  \ln^2  \frac{M^2}{\mu_l^2} - 2 \ln  \frac{Q^2}{\mu_l^2} \ln \frac{M^2}{\mu_l^2} \right]
\end{align}
The first term is the high-scale matching $C$, the second term arises from integrating the $\ln Q^2/\mu^2$ anomalous dimension from $\mu_h$ to $\mu_l$, and the third term is the low-scale matching $D$. The existence of the log term in the matching also follows from the consistency condition that the theory is independent of $\mu_l$. Since changes in the running between $\mu_h$ and $\mu_l$ contain a single log from the anomalous dimension, there must be a single log in the matching. What is non-trivial is that Eq.~(\ref{m}) only requires a single-log in the matching to all orders in perturbation theory~\cite{Chiu:2007yn,Chiu:2009ft}.

The resummed electroweak corrections can be grouped as LL, NLL, etc., in the usual way, and the precise definition for \scetew\ can be found in Ref.~\cite{Chiu:2009mg}. All terms needed for a NLL computation are known, so  \emph{all} processes can be computed to resummed NLL order. Refs.~\cite{Chiu:2009ft,Fuhrer:2010eu}  computed the one-loop $d_S$ and $C$ terms, for all $2 \to 2$ processes. 

The three-loop cusp anomalous dimension $A$ and  two-loop non-cusp anomalous $B$ are known, except for the scalar Higgs contributions, which are numerically small. The two-loop contribution to $D_C$ is not known. The NNLL results are known, with the exception of these terms.

\end{appendix}

\bibliography{EWS}

\end{document}